\documentclass[]{pasj02} 

\usepackage[switch,mathlines]{lineno} 
\usepackage{natbib}

\jyear{2026}
\Received{}
\Accepted{}

\begin{document} 

\title{ Comparing explodability predictions from a parameter-optimized semi-analytic model with structure-based progenitor criteria }

\author{
 Yuxin U. \textsc{Wu},\altaffilmark{1}\altemailmark \email{yuxin@g.ecc.u-tokyo.ac.jp} \orcid{0009-0004-0045-478X}
 Hideyuki \textsc{Umeda},\altaffilmark{1}\orcid{0000-0001-8338-502X}
 and 
 Koh \textsc{Takahashi},\altaffilmark{2} \orcid{0000-0002-6705-6303}
}
\altaffiltext{1}{Department of Astronomy, School of Science, The University of Tokyo, 7-3-1 Hongo, Bunkyo, Tokyo 113-0033, Japan}
\altaffiltext{2}{National Astronomical Observatory of Japan, National Institutes for Natural Science, 2-21-1 Osawa, Mitaka, Tokyo 181-8588, Japan}



\maketitle

\begin{abstract}
Three-dimensional (3D) simulations of neutrino-driven core-collapse supernovae are among the most reliable tools for predicting explosion outcome. However, their high computational cost limits systematic surveys over large progenitor samples. We test how well a fast one-dimensional (1D) approach captures progenitor explodability. We use a parameter-optimized semi-analytic 1D explosion model based on \citet{muller2016}, calibrated to the 3D results of \citet{burrows2024} as an adopted reference set. We compare the model's explodability predictions with commonly used structure-based criteria: compactness, the free-fall mass coordinate, and the two-parameter $\mu_4$--$M_4$ criterion. Our analysis shows that the semi-analytic model can reproduce the trends seen in this adopted 3D calibration set by adjusting physically meaningful parameters. This provides a more direct way to examine the physics that controls explodability than traditional structure-based criteria. We identify where the semi-analytic model agrees with these criteria, where it differs, and which physical trends explain the differences. This work clarifies the strengths and limitations of structure-based explodability criteria by evaluating them against a parameter-optimized, neutrino-driven semi-analytic model.
\end{abstract}


\section{Introduction}

Core-collapse supernovae (CCSNe) are among the most energetic events in the universe. They are key sites of heavy-element production and they form compact remnants.  Despite their importance, the explosion mechanism is still uncertain. Understanding CCSN explosions helps explain neutron-star formation in the Milky Way and constrains the origin of heavy elements. \citep[e.g.,][]{woosley2007b,notomo2013}

Three-dimensional (3D) simulations have improved substantially in recent years. For example, \citet{burrows2024} used the FORNAX code to simulate 20 progenitors in 3D, producing one of the largest 3D CCSN explosion samples to date. \citet{mueller2025} studied low-mass progenitors in 3D and reproduced unusually low-mass neutron stars within the neutrino-driven framework. \citet{nakamura2025} reported that successful explosions correlate with a density drop at the Si/O interface, and all of their models exploded. These results highlight the role of multi-dimensional effects in neutrino-driven explosions.

However, 2D/3D simulations remain computationally expensive, which makes it difficult to cover large progenitor samples and to carry out systematic parameter-space studies. Many efforts that aim to connect supernova theory to observations and to broader astrophysical problems therefore still rely heavily on one-dimensional (1D) frameworks. For example, large-sample studies of explosive nucleosynthesis and yield grids are often built on 1D explosion models and are widely used in nucleosynthesis- and chemical-evolution–related applications \citep[e.g.,][]{sukhbold2016}. Supernova light-curve and spectral modeling also commonly adopts spherically symmetric 1D radiation-hydrodynamics and radiative-transfer calculations as basic tools \citep[e.g.,][]{Morozova2015,Curtis2021}. In this context, fast models based on analytic approximations and calibrated parameterizations remain indispensable: they enable scans over large progenitor sets at an affordable computational cost and provide predictions for explosion outcomes and observables, such as explosion energy, ejecta mass, and $^{56}$Ni yields. \citep[e.g.,][]{oconnor2011,Pejcha2015,Perego2015,sukhbold2018,muller2016}

Previous 1D studies can be grouped into three categories. The first category solves neutrino transport with the Boltzmann equation and follows the evolution from collapse and bounce to shock stagnation or revival \citep[e.g.,][]{oconnor2015, kuroda2016, chan2020}. These calculations are expensive, so multi-dimensional studies often adopt approximate transport schemes to reduce the cost. For example, the isotropic diffusion source approximation \citep[IDSA; e.g.,][]{liebendorfer2009} lowers the computational cost by trading accuracy in the semi-transparent region for the key physical behavior. The leakage scheme \citep[e.g.,][]{ruffert1996} estimates neutrino energy loss using a local diffusion time.

The second category does not model neutrino heating directly. Instead, it triggers explosions by injecting mechanical or thermal energy. The thermal-bomb method \citep[e.g.,][]{shigeyama1988} and the piston-driven method \citep[e.g.,][]{woosley1988} reproduce several observed properties of SN~1987A in 1D. However, these approaches do not provide a clear physical link between progenitor structure and the heating conditions required for explosion.

The third category, which is the focus of this paper, avoids full radiation-hydrodynamic simulations. Instead, it uses analytic scaling relations and ordinary differential equation (ODE) approximations to quickly predict whether a model explodes and, if it does, the explosion outcome \citep{muller2016}. These methods are computationally efficient and can be used to survey large model sets. We adopt the semi-analytic model of \citet{muller2016}. However, its predictive power depends on empirical parameters that encode multi-dimensional effects in an effective way. The original parameter set was calibrated mainly to 2D simulations and thus inherits the calibration biases of 2D dynamics. Multi-dimensional instabilities, including convection and the standing accretion shock instability (SASI), modify the dwell time and turbulent stresses in the gain region, but their net impact differs between 2D and 3D because turbulence and mode coupling are not the same in different dimensions. Recent 3D simulations nevertheless indicate explosions for progenitors that are often difficult to explode in traditional 1D studies. For example, \citet{nakamura2025} and \citet{burrows2024} reported successful explosions for some progenitors in the $22$--$25\,M_\odot$ range. Motivated by these 3D trends, we adjust the empirical parameters of the M\"uller model within physically reasonable ranges and test whether the updated calibration yields improved agreement with 3D outcomes.

In addition to direct simulations, many studies predict explosion success using empirical criteria built from progenitor structure. \citet{oconnor2011} introduced the compactness parameter $\xi_M$ to quickly assess explodability from progenitor structure. The choice of $M$ is not unique and different values mainly reflect the same idea of central concentration; in this work we focus on the commonly used $\xi_{2.5}$. \citet{ertl2016} proposed the two-parameter $\mu_4$--$M_4$ criterion, which is more closely tied to neutrino heating and separates exploding and non-exploding models with an approximately linear boundary. Recent studies continue to test how well such criteria work across progenitor sets. For example, \citet{maltsev2025} suggested that multi-parameter combinations may be required in complicated mass ranges. \citet{luo2025} included rotation and found that it can strongly change explodability for low-metallicity progenitors, and they suggested a higher critical value of $\xi_{2.5}$.

In this paper, we use the semi-analytic neutrino-driven model of \citet{muller2016} as a fast and physically interpretable framework, and we update its empirical parameters by calibrating to recent 3D outcomes \citep{burrows2024}. We therefore use the \cite{burrows2024} results as a practical reference set and explicitly account for the associated theoretical uncertainty in the Discussion and Appendix. Our goals are to test how well a parameter-optimized 1D semi-analytic model reproduces 3D trends in explodability and in explosion properties ($E_{\rm expl}$, $M_{\rm PNS}$, and ejected $^{56}$Ni mass); and to reassess structure-based indicators, including $\xi_{2.5}$, $M_{\rm ff}$, and $\mu_4$--$M_4$, by identifying where they agree with the calibrated model and where they fail. This positioning distinguishes our work from purely structure-criterion studies by using a calibrated neutrino-driven model as the reference while retaining the computational speed needed for large-sample surveys.

The paper is organized as follows. In Section 2 we describe the semi-analytic M\"uller model, the progenitor sets used in this work, and the structural variables ($\xi_{2.5}$, $M_{\rm ff}$, and $\mu_4$--$M_4$) employed for comparison, and our parameter-optimization procedure calibrated to the 3D results of \citet{burrows2024}. In Section 3 we present the optimized-model results and compare explodability and explosion properties with the original calibration and with structure-based criteria. Section 4 discusses the implications for parameter calibration and for the applicability of structure-based indicators. Section 5 summarizes our main conclusions and outlines future directions.

\section{Method}

\subsection{M\"uller method}\label{ssec:21}

\begin{table*}
  \tbl{Adjustable parameters in the M\"uller semi-analytic model.\footnotemark[*]}{%
  \begin{tabular}{ccc}
      \hline
      Parameter & Definition\footnotemark[$\dag$] & Reference \\ 
      \hline
      $\alpha_{\rm turb}$ & Shock expansion factor due to turbulent stresses  & \citet{mueller2015} \\
      $\zeta$ & Efficiency factor for conversion of accretion energy into $\nu$ luminosity  & \citet{mueller2014} \\
      $\tau_{\rm 1.5}$ & Cooling time-scale for a $1.5\,M_\odot$ neutron star & \citet{hudepohl2014} \\
      $\beta_{\rm expl}$ & Shock compression ratio during the explosion phase & \citet{mueller2015.2} \\
      $\alpha_{\rm out}$ & Volume fraction of outflows & \citet{muller2016} \\
      \hline
    \end{tabular}}\label{tab:muller-params}
\begin{tabnote}
\footnotemark[$*$] Five adjustable parameters defined in \citet{muller2016}.  \\ 
\footnotemark[$\dag$] Same as Table 1 in \citet{muller2016}. 
\end{tabnote}
\end{table*}

The M\"uller method is a simplified framework for evaluating the influence of progenitor structure on the explosion outcome. It uses only the pre-collapse progenitor profile as input and combines analytic scaling relations with ordinary differential equations (ODEs) to predict the explosion outcome, including whether an explosion occurs (explodability) and, when it does, the explosion energy, the proto-neutron-star (PNS) mass, and the ejected $^{56}$Ni mass. The model can evaluate a large progenitor set in a single batch within seconds, which makes it suitable for large-scale parameter surveys.

We implemented the full semi-analytic model following \citet{muller2016} and the later implementation details in \citet{takahashi2023}. \citet{takahashi2023} corrected several issues in the original implementation; we adopt their corrected version. Our code reproduces the published results, which provides a baseline validation of the implementation used in this work.

The model assumes that each mass shell falls inward on a timescale proportional to its free-fall time,
\begin{equation}
t = C\,\tau_{\rm ff}(M) = C\sqrt{\frac{\pi}{4G\bar{\rho}}}.
\end{equation}
The time coordinate $t$ is not the physical time. It labels the sequence of shell infall and is used to track how key quantities evolve with the fallback stage. We therefore express all subsequent quantities as functions of this fallback-stage coordinate.

Given the pre-collapse profile, the model evolves $M_{\rm PNS}$ and $E_{\rm diag}$ with the swept-up shock mass $M_{\rm sh}$, while $r_{\rm g}$, $r_{\rm sh}$, and $L_\nu$ are updated from scaling relations. We estimate the gain radius and the shock radius as follows:
\begin{equation}
r_{\rm g} = \sqrt[3]{r_1^3 + \left( \frac{\dot{M}}{M_\odot\,{\rm s}^{-1}} \right)
\left( \frac{M}{M_\odot}\right)^{-3} + r_0^3},
\end{equation}
\begin{align}
r_{\rm sh} &= \alpha_{\rm turb}\times 0.55\,{\rm km}
\left( \frac{L_\nu}{10^{52}\,{\rm erg\,s^{-1}}} \right)^{4/9} \nonumber \\
&\times \left( \frac{M_{\rm PNS}}{M_\odot} \right)^{5/9}
\left( \frac{r_{\rm g}}{10\,{\rm km}} \right)^{16/9} \nonumber \\
&\times \left( \frac{\dot{M}}{M_\odot\,{\rm s}^{-1}}\right)^{-2/3}
\alpha_{\rm redshift}^{6/9}.
\end{align}

The parameter $\alpha_{\rm turb}$ captures the effect of turbulent stresses on the shock radius in an effective 1D form. It approximates multi-dimensional turbulence as a multiplicative correction to the 1D shock-radius scaling. \citet{mueller2015} performed 2D relativistic radiation-hydrodynamic simulations of the 15\,$M_\odot$ progenitor from \citet{woosley2007a}. They imposed initial perturbations with different angular wavenumbers $\ell$ and amplitudes to compare the critical conditions for explosion. They found that explosions occur when the mean squared turbulent Mach number is $\langle \rm{Ma}\rangle^2 \approx 0.3- 0.5$. They related the turbulence level to the shock-expansion factor by assuming $\langle \rm{Ma}^2\rangle=\frac{1}{3}\alpha_{\rm turb}$. \citet{takahashi2023} also revised the exponent of $\alpha_{\rm redshift}$ in the shock-radius scaling from $2/9$ to $6/9$ to reproduce results consistent with \citet{muller2016}.

The model writes the neutrino luminosity as the sum of an accretion term and a diffusion term,
\begin{align}
L_\nu &= L_{\rm acc}+L_{\rm diff} \nonumber\\
&= \zeta\,\frac{GM\dot{M}}{r_{\rm g}}
+ 0.33\,\frac{E_{\rm bind}}{\tau_{\rm cool}}\,\exp\!\left(-\frac{t}{\tau_{\rm cool}}\right),
\end{align}
where $\zeta$ is the efficiency for converting accretion power into neutrino luminosity. \citet{mueller2014} calibrated $\zeta$ by comparing accretion luminosities with the available gravitational energy for several progenitors. We use the corrected form of $L_{\rm diff}$ that includes the factor $1/\tau_{\rm cool}$, as noted by \citet{takahashi2023}.

The cooling timescale is
\begin{equation}
\tau_{\rm cool}=\tau_{1.5}\left(\frac{M}{1.5\,M_\odot}\right)^{5/3},
\end{equation}
where $\tau_{1.5}$ is the cooling timescale for a $1.5\,M_\odot$ proto-neutron star. It is obtained by fitting an exponential decay of the luminosity and therefore requires calibration from numerical simulations.

The model defines an advection timescale and a heating timescale in the gain region,

\begin{align}
\tau_{\rm adv} &= 18\,{\rm ms}\,
\left(\frac{r_{\rm sh}}{100\,{\rm km}}\right)^{3/2}
\left(\frac{M}{M_\odot}\right)^{-1/2}
\ln\!\left(\frac{r_{\rm sh}}{r_{\rm g}}\right),
\end{align}
\begin{align}
\tau_{\rm heat} &= 150\,{\rm ms}\,
\left(\frac{e_g}{10^{19}\,{\rm erg}}\right)
\left(\frac{r_{\rm g}}{100\,{\rm km}}\right)^2
\left(\frac{L_\nu}{10^{52}\,{\rm erg\,s^{-1}}}\right)^{-1} \\
&\times \left(\frac{M}{M_\odot}\right)^{-2}
\alpha_{\rm redshift}^{-3/2}.
\end{align}
The timescale $\tau_{\rm adv}$ measures how long matter resides in the gain region and can absorb neutrino heating. The timescale $\tau_{\rm heat}$ measures how quickly neutrino heating can supply enough energy to overcome gravitational binding. The evolution is split into a pre-explosion phase and an explosion phase, with the transition defined by
\begin{equation}
\frac{\tau_{\rm adv}}{\tau_{\rm heat}} > 1.
\end{equation}

During the explosion phase, newly shocked matter is heated and accelerated outward. In the early stage, the post-shock velocity $v_{\rm post}$ can remain below the escape velocity $v_{\rm esc}$, so the flow stays gravitationally bound and accretion can continue. \citet{muller2016} therefore divides the explosion phase into two stages (phase I and phase II) based on whether $v_{\rm post}$ exceeds $v_{\rm esc}$.

The post-shock velocity is
\begin{align}
v_{\rm post} =&\ 0.794\,
\frac{\beta_{\rm expl}-1}{\beta_{\rm expl}}
\left(\frac{E_{\rm diag}}{M-M_{\rm ini}}\right)^{1/2}
\left(\frac{M-M_{\rm ini}}{\rho r^3}\right).
\end{align}
The diagnostic explosion energy evolves with the swept-up shock mass as
\begin{align}
\frac{{\rm d}E_{\rm diag}}{{\rm d}M_{\rm sh}} =&\
\frac{\epsilon_{\rm rec}\eta_{\rm acc}}{e_g}
\min\!\left(1,\frac{\dot{M}}{4\pi r^2 v_{\rm sh}\rho}\right)
+ \alpha_{\rm out}\,(\epsilon_{\rm bind}+\epsilon_{\rm burn}).
\end{align}
The first term represents an indirect energy supply associated with accretion-enhanced neutrino heating, and the second term represents the direct contribution from shock-heated outflow, including binding-energy changes and nuclear-burning release. After the evolution enters phase II, the model treats accretion as negligible and drops the indirect accretion-related contribution.

The proto-neutron-star mass is evolved in the same two stages. In phase I,
\begin{align}
\frac{{\rm d}M_{\rm PNS}}{{\rm d}M_{\rm sh}}=
(1-\alpha_{\rm out})
\left(1-\frac{\eta_{\rm acc}}{|e_g|}\right),
\end{align}
and in phase II,
\begin{align}
\frac{{\rm d}M_{\rm PNS}}{{\rm d}M_{\rm sh}}=0.
\end{align}

Two parameters, $\beta_{\rm expl}$ and $\alpha_{\rm out}$, enter the expressions for $v_{\rm post}$, $E_{\rm diag}$, and $M_{\rm PNS}$. The parameter $\beta_{\rm expl}$ is the compression ratio across the shock. A radiation-dominated strong shock gives $\beta_{\rm expl}\approx 7$ for $\gamma=4/3$, while \citet{mueller2015.2} found values closer to $\beta_{\rm expl}\approx 4$ in simulations. Effects that can reduce the average compression include weaker shocks during the explosion phase, nuclear burning, additional pressure components in the post-shock region, and departures from spherical symmetry. The parameter $\alpha_{\rm out}$ sets the fraction of shocked material that becomes neutrino-driven outflow.

We classify two types of explosion failure. Models that never satisfy $\tau_{\rm adv}/\tau_{\rm heat}>1$ are labeled Implosion (non-revival). Models that satisfy $\tau_{\rm adv}/\tau_{\rm heat}>1$ but later fail because the neutron star exceeds its maximum mass are labeled Implosion (post-revival). We adopt a gravitational-mass limit of $M_{\rm grav}=2.05\,M_\odot$ for neutron stars \citep[e.g.,][]{antoniadis2013, demorest2010}. The baryonic mass is converted from the gravitational mass using
\begin{align}
M_{\rm PNS}=M_{\rm grav}+0.084\left(\frac{M_{\rm grav}}{M_\odot}\right)^2 M_\odot,
\end{align}
which gives a maximum baryonic mass of $M_{\rm PNS}=2.403\,M_\odot$ for this limit.

This section summarizes the key equations of \citet{muller2016}. The full computational procedure is described in \citet{muller2016} and Appendix~A of \citet{takahashi2023}, so we do not repeat the derivations here. This work focuses on the five adjustable parameters listed in Table~\ref{tab:muller-params}. We keep their definitions and typical ranges consistent with \citet{muller2016} and update them when comparing to recent multi-dimensional results.

Finally, \citet{takahashi2023} proposed the free-fall mass coordinate $M_{\rm ff}$ to characterize the pre-supernova structure. We discuss $M_{\rm ff}$ in Section~2.3.

\subsection{Progenitor sets}

This work primarily compares our semi-analytic predictions with the 3D explosion results summarized by \citet{burrows2024}. Their FORNAX simulations were carried out for progenitors from \citet{sukhbold2016, sukhbold2018}. This dataset is among the most widely used and best-sampled progenitor suites in recent multi-dimensional CCSN studies. It has also been adopted in other large surveys, including the 2D simulations of \citet{vartanyan2023} and the 3D simulations of \citet{nakamura2025}. Throughout this paper, we refer to \citet{sukhbold2016} as S16 and \citet{sukhbold2018} as S18. We note, however, that these progenitor models (as for all stellar-evolution calculations) are not free from uncertainties. In particular, the treatment of convective boundaries and the numerical resolution during core-helium burning can affect the resulting pre-SN core structure and may introduce stochastic variations in quantities that are relevant for explodability \citep[e.g.,][]{Chieffi2020}. Our conclusions should therefore be interpreted as applying to these commonly used progenitor sets as practical testbeds, rather than as definitive statements about the explodability of all realistic massive-star progenitors.

The progenitor set used by \citet{muller2016} (hereafter M16) was computed with the KEPLER code and is close to S16 in its overall setup. It differs mainly in its microphysics: it adopts the neutrino energy-loss rates of \citet{itoh1996} and the solar-abundance scale of \citet{asplund2009}. These choices shift several structural features, including the carbon-burning mode and the locations of shell interfaces, toward lower ZAMS masses. The shift is $\sim 1.5\,M_\odot$.

\begin{figure}
 \begin{center}
  \includegraphics[width=8cm]{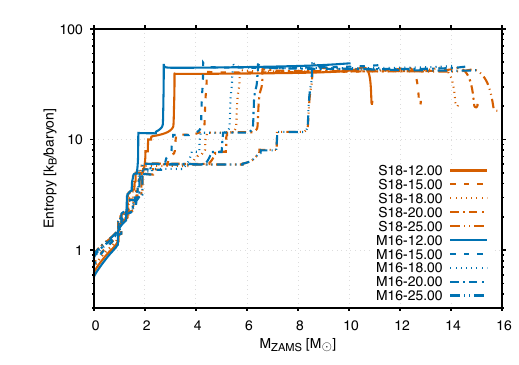} 
  \includegraphics[width=8cm]{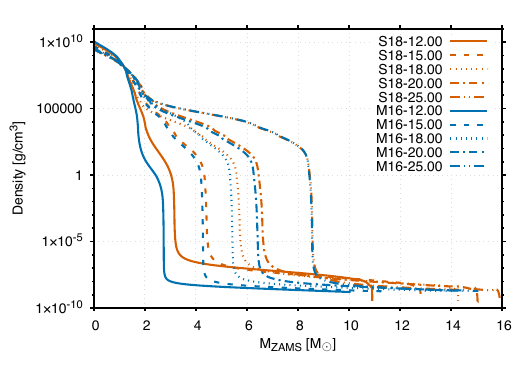}
 \end{center}
\caption{Comparison of entropy and density profiles between S18 and M16 at the same ZAMS masses. We show models with ZAMS masses of 12, 15, 18, 20, and $25\,M_\odot$. Because S18 corrected an error in pair-neutrino energy losses in the older KEPLER code, S18 models show higher entropy and lower density overall. \\
Alt text: Two vertically stacked line plots comparing entropy and density profiles for S18 and M16 at five ZAMS masses. S18 curves generally show higher entropy and lower density than M16.}
\label{fig:progenitor-entropy-density}
\end{figure}

S18 incorporates the same microphysical updates (the \citet{itoh1996} cooling rates and the \citet{asplund2009} abundance scale), so its inputs are more consistent with M16. A key additional change in S18 is a fix to a pair-neutrino-related coding error in KEPLER, which had underestimated the energy-loss rate by about a factor of two. Figure~\ref{fig:progenitor-entropy-density} compares S18 and M16 models at the same ZAMS masses and shows that S18 has higher entropy and lower density overall, consistent with stronger neutrino cooling. S18 also provides higher interior resolution, reaching a mass resolution of $\sim 10^{-3}\,M_\odot$ in the core. Because structural features such as the entropy gradient at the Si/O interface can strongly affect explodability \citep[e.g.,][]{summa2016, boccioli2023}, this higher resolution helps make structure-based diagnostics more reliable.

Finally, when comparing different progenitor sets, trends shown purely as a function of $M_{\rm ZAMS}$ should be interpreted with caution, because differences in microphysics and stellar-evolution assumptions can change the mapping between initial mass and the final core structure. A more physical comparison coordinate is often the CO-core mass $M_{\rm CO}$, which summarizes the outcome of late-stage burning and is less sensitive to the initial-mass mapping \citep[e.g.,][]{Patton2020}. For this reason, in addition to presenting results versus $M_{\rm ZAMS}$ for continuity with earlier work, we also examine explodability and explosion properties as functions of $M_{\rm CO}$ in our Results section.

Table~\ref{tab:progenitor-sets} summarizes the mass coverage and the minimum mass spacing for S16, S18, and M16.

\begin{table*}
  \tbl{Progenitor sets.\footnotemark[$*$]}{%
  \begin{tabular}{cccc}
      \hline
      Progenitor set & Reference & Mass range [$M_\odot$] & Minimum mass spacing [$M_\odot$] \\ 
      \hline
      S16 & \cite{sukhbold2016} & 9--120 & 0.1 \\
      S18 & \cite{sukhbold2018} & 12--27 & 0.01 \\
      M16 & \cite{muller2016} & 9.5--45 & 0.01 \\
      \hline
    \end{tabular}}\label{tab:progenitor-sets}
\begin{tabnote}
\footnotemark[$*$] Three progenitor sets with different mass ranges and mass spacing. \\
\end{tabnote} 
\end{table*}

\subsection{Structure variables}

We compute several commonly used structural parameters that link pre-supernova progenitor structure to explodability. Many previous studies have summarized such quantities based on one-dimensional hydrodynamic calculations. In our sample, three parameters show the clearest correlation with explodability: the compactness $\xi_{2.5}$, the free-fall mass coordinate $M_{\rm ff}$, and the two-parameter $\mu_4$--$M_4$ criterion. We therefore focus on these three in the following subsections.

\subsubsection{compactness}

\citet{oconnor2011} defined the compactness parameter
\begin{equation}
\xi_M=\frac{M/M_\odot}{r(M)/1000\,{\rm km}}.
\end{equation}
They recommended evaluating $\xi_M$ at core bounce. Later work showed that using the pre-collapse progenitor profile yields similar discriminative power. \citet{ugliano2012} found that computing $\xi_M$ directly from the progenitor structure does not significantly change its ability to separate exploding and non-exploding cases. \citet{sukhbold2014} also reported that values evaluated when the infall velocity reaches $1000\,{\rm km\,s^{-1}}$ are close to those at core bounce.

\begin{figure}
 \begin{center}
  \includegraphics[width=8cm]{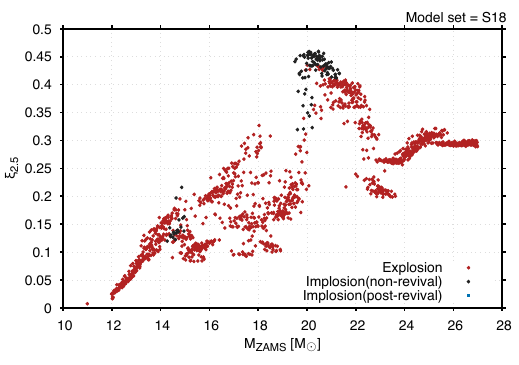}
  \includegraphics[width=8cm]{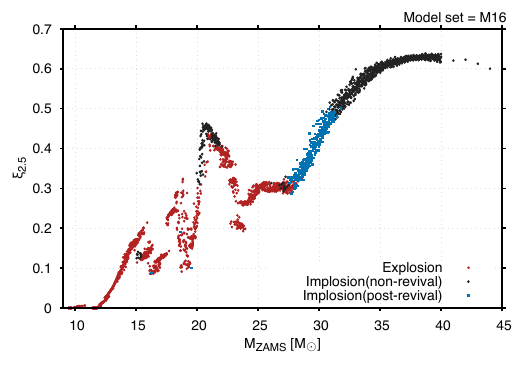}
 \end{center}
\caption{Compactness $\xi_{2.5}$ for S18 (top) and M16 (bottom). Colors indicate the outcome predicted by the semi-analytic model: red, successful explosions; black, failures that never satisfy the shock-revival condition; blue, cases that satisfy the shock-revival condition but later collapse to a black hole because the neutron-star mass exceeds the adopted limit. The M16 panel is consistent with Figure~27 of \citet{takahashi2023}.\\
Alt text: Two vertically stacked scatter plots of compactness versus ZAMS mass for S18 and M16. Red, black, and blue points mark explosion, non-revival implosion, and post-revival implosion, with failures clustering near local peaks.}
\label{fig:structure-compactness}
\end{figure}

The M\"uller method used in this work does not follow the collapse phase to core bounce, so we compute $\xi_M$ from the pre-collapse progenitor profiles. Figure~\ref{fig:structure-compactness} shows $\xi_{2.5}$ for S18 (top) and M16 (bottom). Our M16 results are consistent with Figure~27 of \citet{takahashi2023} and broadly agree with the corresponding trends shown in \citet{muller2016}. In our sample, $\xi_{2.5}$ correlates with explodability: models in high-compactness regions are more often classified as failed explosions by the semi-analytic model.

\subsubsection{Free-fall mass coordinate}

\begin{figure}
 \begin{center}
  \includegraphics[width=0.48\linewidth]{ 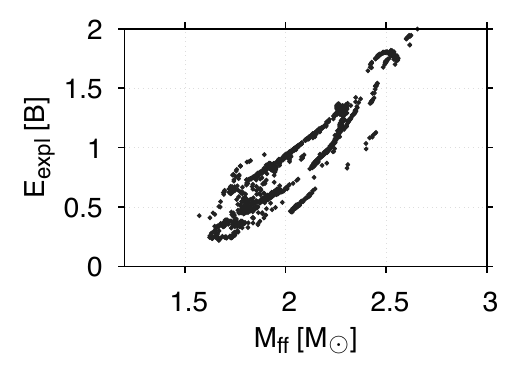}
  \includegraphics[width=0.48\linewidth]{ 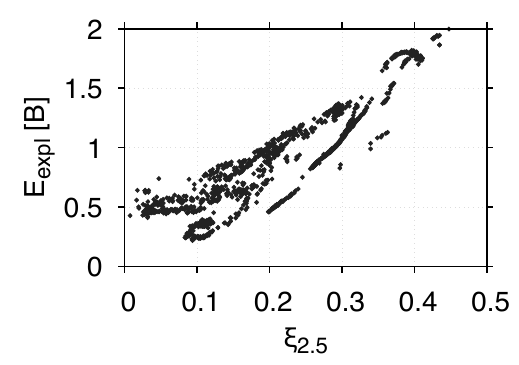}
  \includegraphics[width=0.48\linewidth]{ 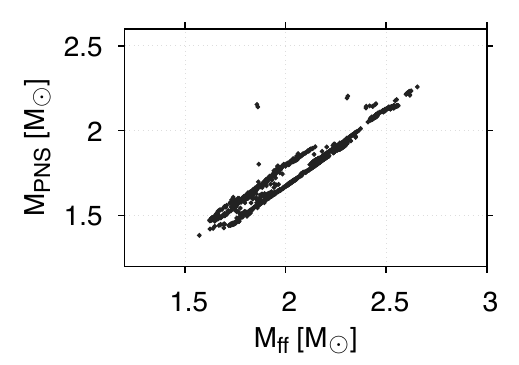}
  \includegraphics[width=0.48\linewidth]{ 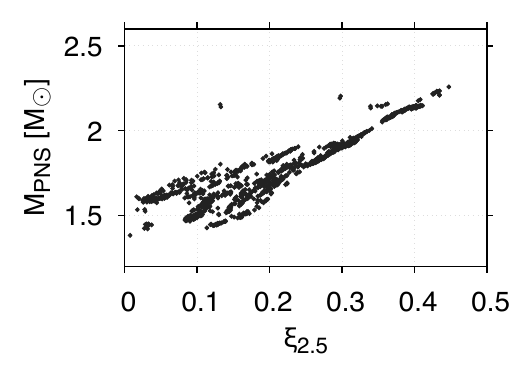}
  \includegraphics[width=0.48\linewidth]{ 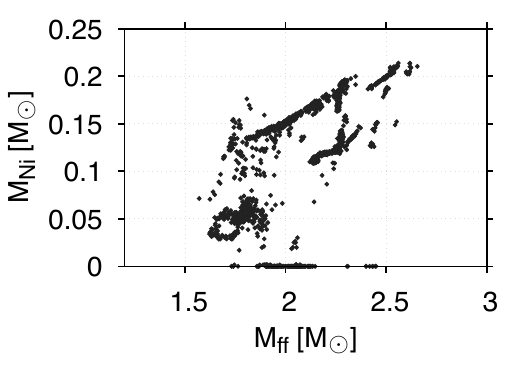}
  \includegraphics[width=0.48\linewidth]{ 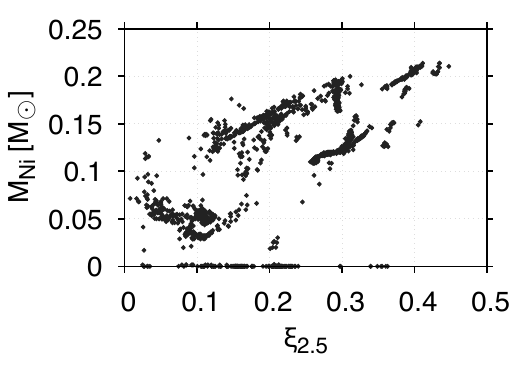}
 \end{center}
\caption{Explosion characteristics as functions of $M_{\rm ff}$ (left) and $\xi_{2.5}$ (right) for S18. From top to bottom: explosion energy, PNS mass, and ejected $^{56}$Ni mass. In these examples, the trends are more nearly monotonic with $M_{\rm ff}$ than with $\xi_{2.5}$, especially for the PNS mass (see \citet{takahashi2023}). \\
Alt text: Six scatter plots arranged in three rows and two columns for S18. Explosion energy, proto-neutron-star mass, and ejected nickel mass are plotted against free-fall mass coordinate on the left and compactness on the right; trends are smoother and more nearly monotonic with free-fall mass coordinate.}
\label{fig:structure-monotonic}
\end{figure}

\citet{takahashi2023} introduced the free-fall mass coordinate $M_{\rm ff}$, defined as the enclosed mass on an iso--free-fall-time surface. They showed that many properties of the progenitor and the explosion, including density, entropy, composition, evolutionary timescales, and explosion characteristics, often vary monotonically with $M_{\rm ff}$. The PNS mass shows an especially clear monotonic trend with $M_{\rm ff}$ (Figure~\ref{fig:structure-monotonic}).

\begin{figure}
 \begin{center}
  \includegraphics[width=8cm]{ 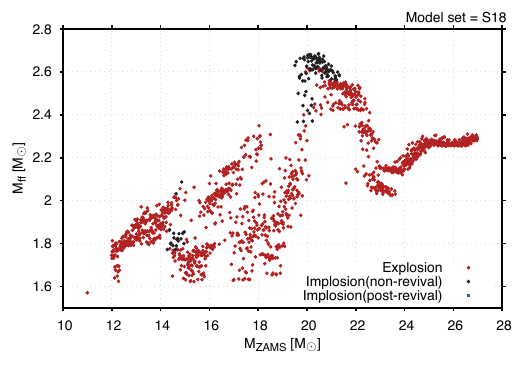}
  \includegraphics[width=8cm]{ 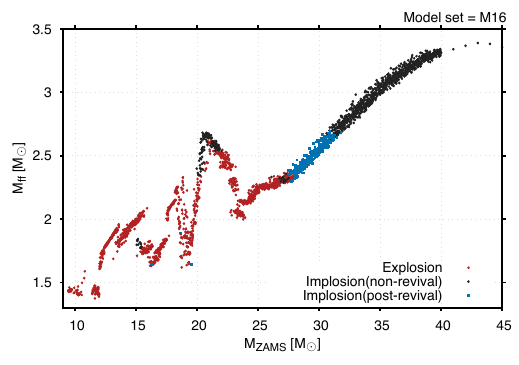}
 \end{center}
\caption{Free-fall mass coordinate $M_{\rm ff}$ for S18 (top) and M16 (bottom). Color coding is the same as in Figure~\ref{fig:structure-compactness}. \\
Alt text: Two vertically stacked scatter plots of free-fall mass coordinate versus ZAMS mass for S18 and M16. Red, black, and blue points mark explosion, non-revival implosion, and post-revival implosion, with failures clustering near local peaks.}
\label{fig:structure-mff}
\end{figure}

Figure~\ref{fig:structure-mff} shows $M_{\rm ff}$ as a function of ZAMS mass together with the explodability classification from the semi-analytic model. The overall trend of $M_{\rm ff}$ closely follows that of the compactness $\xi_{2.5}$, consistent with both quantities tracing core structure. In our sample, models near the peak of these curves are more often classified as failed explosions by the semi-analytic model.

\subsubsection{$\mu_4$--$M_4$ two-parameter criterion}

Single-parameter indicators such as $\xi_{2.5}$ and $M_{\rm ff}$ do not cleanly separate the mass range where exploding and non-exploding models overlap. Several studies therefore argue that combining two or more structural parameters can improve discrimination \citep[e.g.,][]{maltsev2025}. One widely used example is the two-parameter $\mu_4$--$M_4$ criterion of \citet{ertl2016}. It differs from compactness-type measures by using (i) the enclosed mass at the entropy surface $s=4$,
\begin{align}
M_4 \equiv \frac{m(s=4)}{M_\odot},
\end{align}
and (ii) the normalized mass gradient at the same location,
\begin{align}
\mu_4 \equiv \left.\frac{{\rm d}(m/M_\odot)}{{\rm d}\left[r/(1000\,{\rm km})\right]}\right|_{s=4}.
\end{align}
These quantities are connected to the mass accretion rate during the shock-stagnation phase and the neutrino luminosity, and they can therefore be used to assess explodability.

We follow \citet{ertl2016} and evaluate $\mu_4$ with a finite difference using $\delta m=0.3\,M_\odot$,
\begin{align}
\mu_4 &=
\frac{\left[(M_4+\delta m)/M_\odot\right]-M_4}
{\left[r(M_4+\delta m)-r(s=4)\right]/(1000\,{\rm km})} \nonumber\\
&= \frac{0.3}
{\left[r(M_4+0.3\,M_\odot)-r(s=4)\right]/(1000\,{\rm km})}.
\end{align}
Compared with $\xi_{2.5}$ and $M_{\rm ff}$, the $\mu_4$--$M_4$ plane more clearly separates the branch of imploding models that deviates from the main trend. These progenitors can overlap with exploding models in one-parameter representations, so the two-parameter criterion helps reduce this ambiguity.

\begin{figure}
 \begin{center}
  \includegraphics[width=8cm]{ 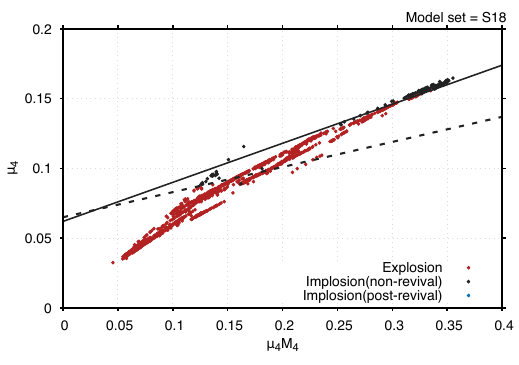}
  \includegraphics[width=4cm]{ 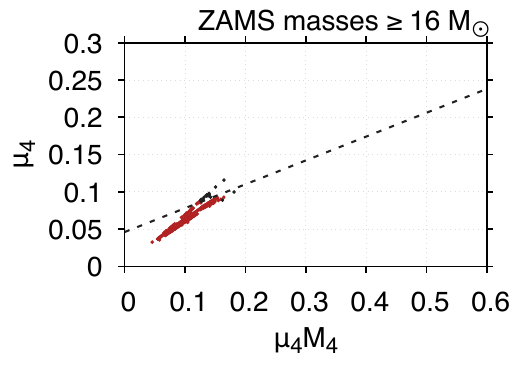}
  \includegraphics[width=4cm]{ 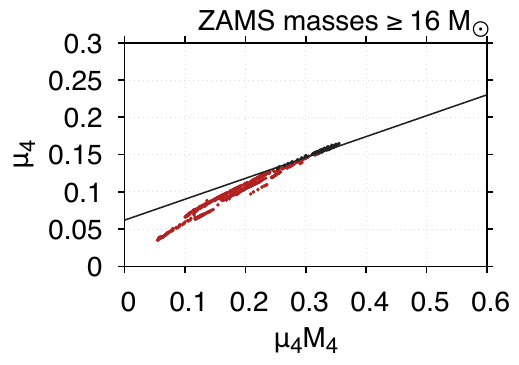}
  \includegraphics[width=8cm]{ 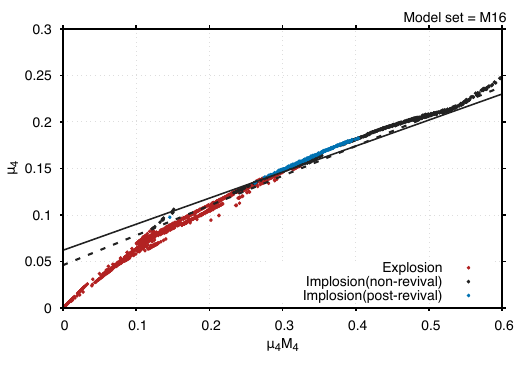}
  \includegraphics[width=4cm]{ 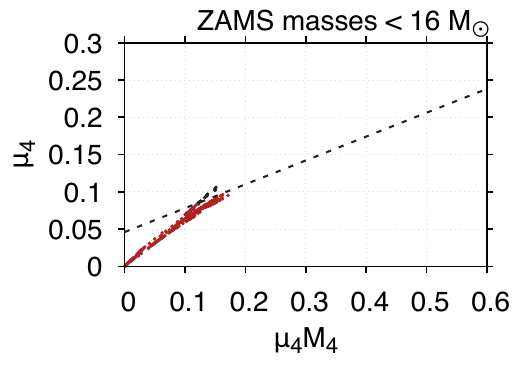}
  \includegraphics[width=4cm]{ 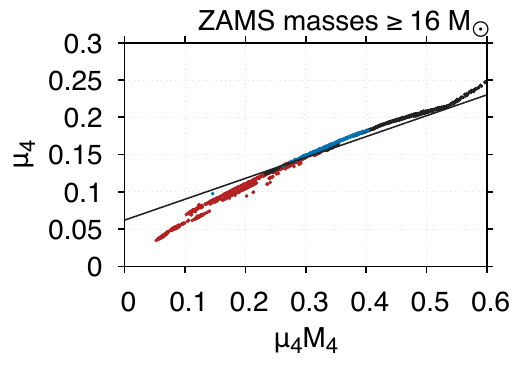}
 \end{center}
\caption{$\mu_4$--$M_4$ results for S18 (top) and M16 (bottom), computed following \citet{ertl2016}. Color coding is the same as in Figure~\ref{fig:structure-compactness}. The solid black line shows the fitted explodability boundary for progenitors with ZAMS masses $\ge16\,M_\odot$, and the dashed black line shows the boundary for ZAMS masses $<16\,M_\odot$. \\
Alt text: A six-panel figure showing the $\mu_4$--$M_4$ relation for S18 and M16. Large upper panels and smaller lower panels display red, black, and blue points with fitted black boundary lines, showing that the two-parameter plane separates exploding and imploding branches more clearly than one-parameter plots.}
\label{fig:structure-mu4M4}
\end{figure}

In our sample, using separate linear boundaries for low- and high-mass progenitors improves the classification. We therefore fit two boundaries, split at $16\,M_\odot$. We define $x\equiv \mu_4 M_4$ and fit a line of the form $\mu_4 = kx+b$. Figure~\ref{fig:structure-mu4M4} shows the results. For M16, the best-fit boundary is $\mu_4 = 0.32\,(\mu_4 M_4)+0.046$ for progenitors with ZAMS masses $<16\,M_\odot$, with 0/493 misclassified. For ZAMS masses $\ge16\,M_\odot$, the boundary is $\mu_4 = 0.28\,(\mu_4 M_4)+0.020$, with 46/2294 misclassified. For S18, the boundary is $\mu_4 = 0.18\,(\mu_4 M_4)+0.065$ for ZAMS masses $<16\,M_\odot$, with 1/400 misclassified, and $\mu_4 = 0.28\,(\mu_4 M_4)+0.062$ for ZAMS masses $\ge16\,M_\odot$, with 8/1100 misclassified.

\subsection{Optimization work}

Figures~\ref{fig:structure-compactness} and \ref{fig:structure-mff} show that, in the ZAMS-mass range $\sim 19$-$21\,M_\odot$, the M\"uller model classifies most progenitors as failed explosions. A similar difficult-to-explode region was also reported by \citet{sukhbold2016} using the central energy-injection approach. Recent 3D studies based on the Sukhbold progenitor set, however, indicate that some progenitors that are difficult to explode in 1D can still explode in 3D. Because 3D simulations are computationally expensive, it is useful to use available 3D outcomes to constrain and update faster 1D models. This strategy reduces the mismatch between 1D and 3D predictions while preserving the ability of 1D frameworks to scan large progenitor samples. In this section, we optimize the parameters of the M\"uller model using 3D results as constraints.

\begin{figure}
    \begin{center}
    \includegraphics[width=0.45\linewidth]{ 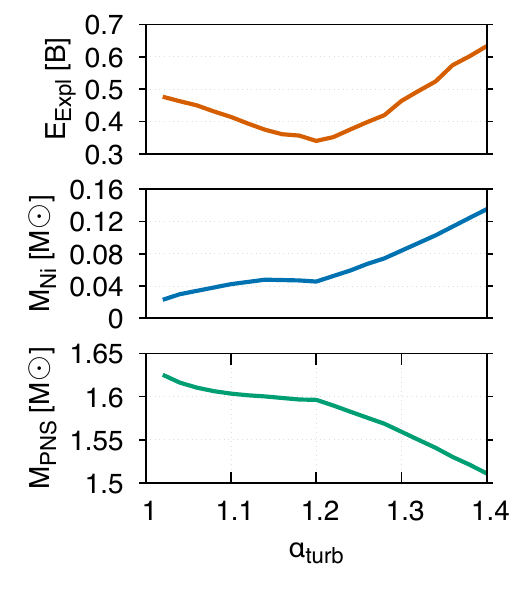}
    \includegraphics[width=0.45\linewidth]{ 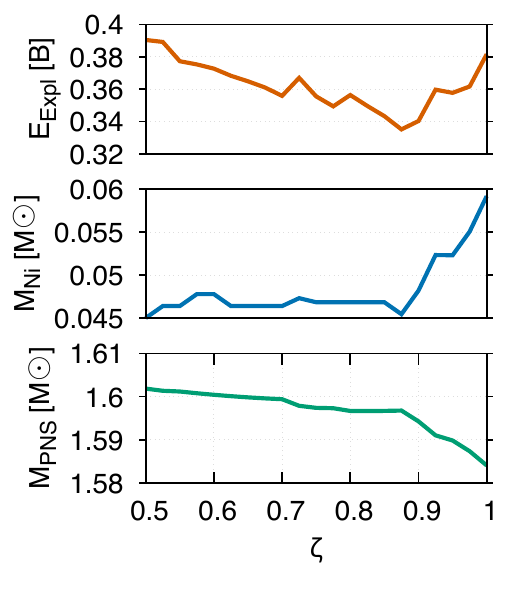}
    \includegraphics[width=0.45\linewidth]{ 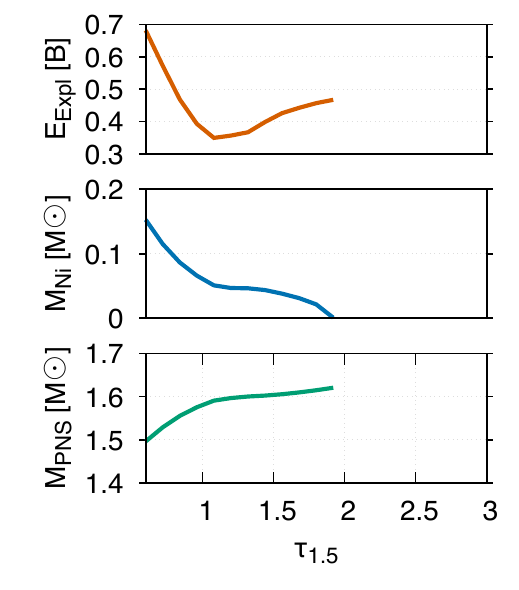}
    \includegraphics[width=0.45\linewidth]{ 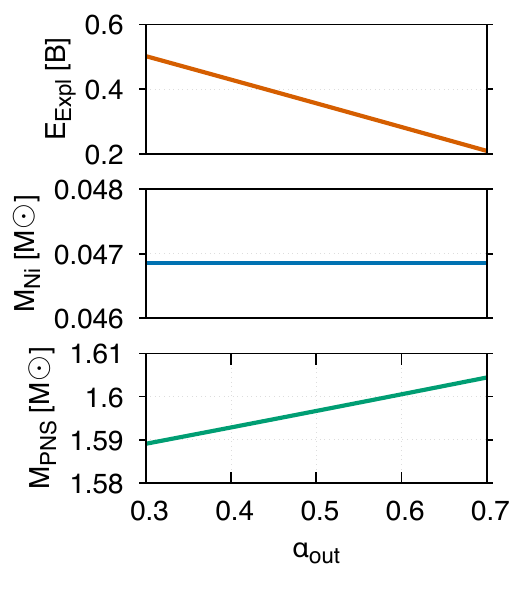}
    \includegraphics[width=0.45\linewidth]{ 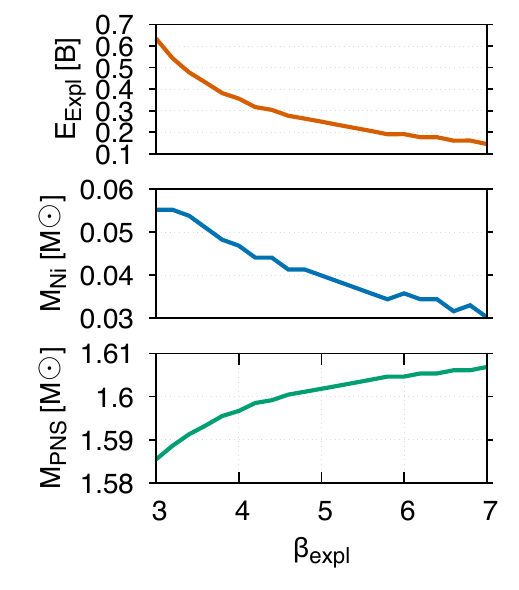}
    \end{center}
    \caption{Sensitivity of the predicted explosion outcome to each parameter for a $16\,M_\odot$ progenitor. In each panel, we vary one parameter while fixing the other four to the standard values of \citet{muller2016} (Table~\ref{tab:opt-values}). \\
    Alt text: Five groups of small line plots showing how predicted explosion energy, nickel mass, and proto-neutron-star mass vary as each model parameter is changed for a $16\,M_\odot$ progenitor. Different parameters affect the three outcomes in different ways.}
    \label{fig:optimization-prmt}
\end{figure}

The M\"uller semi-analytic model contains five adjustable parameters that represent multi-dimensional effects within a 1D framework, but their prior values are not uniquely determined. In the original series of papers, these parameters were mainly calibrated against 2D simulations, so further updates are possible. We therefore recalibrate the parameters by minimizing the discrepancy between 1D explosion outcomes and 3D simulation results. Figure~\ref{fig:optimization-prmt} illustrates how each parameter affects the predicted outcome for a $16\,M_\odot$ progenitor.

We scan the five model parameters within the ranges given by \citet{muller2016}. We define the best-fit set as the one that minimizes the mismatch between the 1D predictions and the 3D results of \citet{burrows2024} for three explosion-outcome quantities: the explosion energy, the PNS mass, and the ejected $^{56}$Ni mass. We quantify the mismatch with a $\chi^2$-like mismatch metric,
\begin{align}
\chi_k^2 &=
\left(\frac{E_{{\rm 1D},k}-E_{{\rm 3D},k}}{E_{{\rm 3D},k}}\right)^2\\
&+\left(\frac{M_{{\rm PNS,1D},k}-M_{{\rm PNS,3D},k}}{M_{{\rm PNS,3D},k}}\right)^2\\
&+\left(\frac{M_{{\rm Ni,1D},k}-M_{{\rm Ni,3D},k}}{M_{{\rm Ni,3D},k}}\right)^2,
\end{align}
and its sample average,
\begin{align}
\bar{\chi}_{\rm tot}^2=\frac{1}{N}\sum_{k=1}^{N}\chi_k^2,
\end{align}
where $k$ labels the progenitor model and $N$ is the number of exploding models included in the comparison. Smaller $\bar{\chi}_{\rm tot}^2$ indicates better agreement between the 1D and 3D outcome.

In the FORNAX 3D survey of \citet{burrows2024}, 15 progenitors exploded (S16: 9.0, 9.25, 9.5, 11, and $60\,M_\odot$; S18: 15.01, 16, 17, 18, 18.5, 19, 20, 23, 24, and $25\,M_\odot$). 
Two of these models (19.56 and $40\,M_\odot$) later collapsed to black holes after shock revival. Because the M\"uller model does not describe this late-time evolution, we treat these cases as successful explosions when forming the calibration sample, performing the same processing as with the successfully exploded sample.

Among the five parameters, we treat $\alpha_{\rm turb}$ differently because it is the shock-expansion factor that accounts for turbulent stresses in an effective 1D manner \citep{mueller2015}. In multi-dimensional simulations, the turbulent Mach number in the gain region depends on the balance between neutrino heating and accretion, and it therefore varies systematically with progenitor structure. This motivates allowing $\alpha_{\rm turb}$ to vary across progenitors, rather than adopting a single constant value. Related discussion is provided in Section 4.2. 

\begin{figure}
    \begin{center}
    \includegraphics[width=1\linewidth]{ 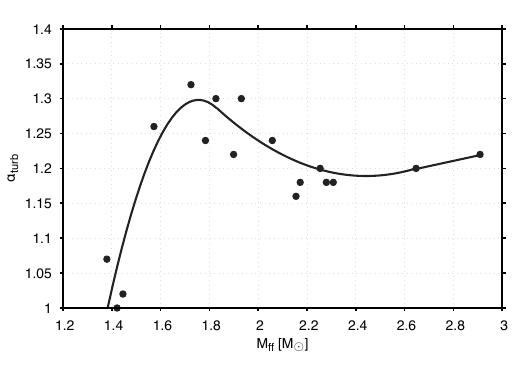}
    \end{center}
    \caption{Best-fit $\alpha_{\rm turb}$ values (black points) and the fitted piecewise relation as a function of $M_{\rm ff}$ (black line). \\
    Alt text: Scatter plot of best-fit $\alpha_{\rm turb}$ versus free-fall mass coordinate, with black points and a fitted black curve. $\alpha_{\rm turb}$ rises at low values, declines at intermediate values, and becomes nearly flat at high values.
}\label{fig:optimization-fitaturb}
\end{figure}

We optimize $\alpha_{\rm turb}$ first. We fix the other four parameters to the standard values of \citet{muller2016} and scan $\alpha_{\rm turb}$ for each progenitor to minimize $\chi_k^2$ (with $N=1$ for each single-model fit). We then fit the resulting best-fit $\alpha_{\rm turb}$ as a function of $M_{\rm ff}$. Figure~\ref{fig:optimization-fitaturb} shows the best-fit $\alpha_{\rm turb}$ values and the fitted trend. The fitted relation is
\begin{equation}
\alpha_{\mathrm{turb}}=
\begin{cases}
-2.15\,M_{\mathrm{ff}}^{2}+7.55\,M_{\mathrm{ff}}-5.33, & M_{\mathrm{ff}}\le 1.830,\\
0.26\,M_{\mathrm{ff}}^{2}-1.27\,M_{\mathrm{ff}}+2.74, & 1.830< M_{\mathrm{ff}}\le 2.647,\\
0.076\,M_{\mathrm{ff}}+0.998, & M_{\mathrm{ff}}>2.647.
\end{cases}
\label{eq:alpha_turb_fit}
\end{equation}

The parameters $\alpha_{\rm turb}$, $\zeta$, and $\tau_{1.5}$ enter the timescale ratio $\tau_{\rm adv}/\tau_{\rm heat}$ and therefore directly affect the explodability criterion. We first fix $\alpha_{\rm turb}$ to the fitted function $\alpha_{\rm turb}(M_{\rm ff})$ and then perform a two-dimensional scan over $\zeta$ and $\tau_{1.5}$. This scan yields best-fit values of $\zeta=0.9$ and $\tau_{1.5}=1.44$. Next, we fix $\alpha_{\rm turb}$, $\zeta$, and $\tau_{1.5}$ and scan the remaining two parameters, $\beta_{\rm expl}$ and $\alpha_{\rm out}$. This step yields $\beta_{\rm expl}=7$ and $\alpha_{\rm out}=0.32$. Table~\ref{tab:opt-values} summarizes the original parameter values from \citet{muller2016} and the optimized values obtained in this work.

\begin{table*}
  \tbl{Optimized parameter values.\footnotemark[$*$]}{%
  \begin{tabular}{cccc}
      \hline
      Parameter & Typical range & Original value & Optimized value \\ 
      \hline
      $\alpha_{\rm turb}$ & 1--1.4 & 1.18 & Piecewise functions related to $M_{\rm ff}$\footnotemark[$\dag$] \\
      $\zeta$ & 0.5--1 & 0.8 & 0.9 \\
      $\tau_{1.5}$ & 0.6--3 & 1.2 & 1.44 \\
      $\alpha_{\rm out}$ & 0.3--0.7 & 0.5 & 0.32 \\
      $\beta_{\rm expl}$ & 3--7 & 4 & 7 \\
      \hline
    \end{tabular}}
\label{tab:opt-values}
\begin{tabnote}
\footnotemark[$*$] Typical ranges and original values are from \citet{muller2016}. Optimized values are obtained from the best-fit procedure described in this section. \\
\footnotemark[$\dag$] The specific functional form is given in equation~(\ref{eq:alpha_turb_fit}).
\end{tabnote}
\end{table*}

\section{Results}

\begin{figure*}
    \begin{center}
    \includegraphics[width=0.7\linewidth]{ 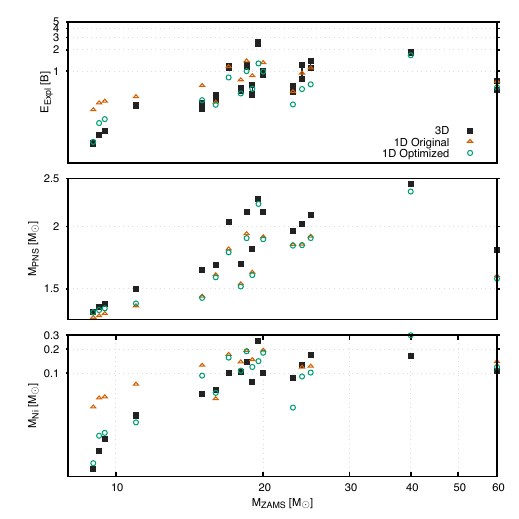}
    \end{center}
    \caption{Explosion outcome predicted with the updated parameter set (green circles), compared with the original parameter set (orange triangles) and the 3D results (black squares). The updated parameters improve the agreement with 3D outcome, and they remove several mismatches in the explosion classification between the original 1D model and the 3D results. \\
    Alt text: Three vertically stacked comparison plots of explosion energy, proto-neutron-star mass, and nickel mass versus ZAMS mass. Black squares show 3D results, orange triangles show the original 1D model, and green circles show the optimized 1D model; the optimized model lies closer to the 3D points overall.}
    \label{fig:results:comparison}
\end{figure*}

Figure~\ref{fig:results:comparison} shows that the updated parameter set improves the agreement with the 3D outcome compared to the original parameters. Quantitatively, the mean mismatch $\bar{\chi}_{\rm tot}^2$ decreases from [4.2] to [0.38] for the calibration sample. Since 19.56 and $40\,M_\odot$ were classified as non-revival implosion during the original parameter set calculation and thus excluded from the fit metric computation for original parameter sets. 

The improvement is strongest for low-mass progenitors ($M<10\,M_\odot$): the predicted explosion energy $E_{\rm expl}$, ejected $^{56}$Ni mass $M_{\rm Ni}$, and PNS mass $M_{\rm PNS}$ move closer to the 3D values. The $19.56\,M_\odot$ and $40\,M_\odot$ progenitors also change outcome. Under the original parameters, they fail to explode and collapse to black holes without shock revival. With the updated parameters, they achieve shock revival and are classified as successful explosions, consistent with the trend reported by \citet{burrows2024}. Overall, the updated parameters shift the model outcome toward the 3D results.

The updated parameters differ systematically from the original set. The best-fit values of $\zeta$, $\tau_{1.5}$, and $\beta_{\rm expl}$ increase, while $\alpha_{\rm out}$ decreases. The fitted $\alpha_{\rm turb}(M_{\rm ff})$ rises with $M_{\rm ff}$ at low $M_{\rm ff}$, decreases at intermediate $M_{\rm ff}$, and approaches a nearly constant value at high $M_{\rm ff}$. Because the calibration sample is limited, this functional form is not unique. Alternative fits produce different trends. We discuss the physical implications and the sensitivity to the fitting choice in Section 4.2.

\begin{figure*}
    \begin{center}
    \includegraphics[width=0.45\linewidth]{ 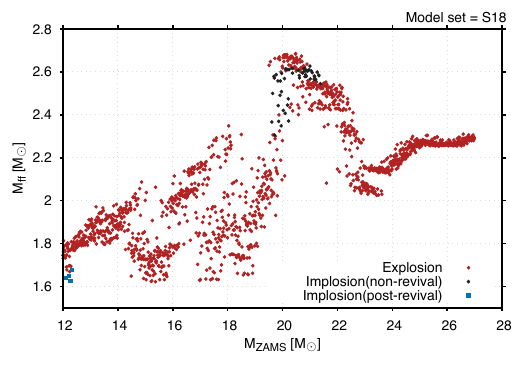}
    \includegraphics[width=0.45\linewidth]{ 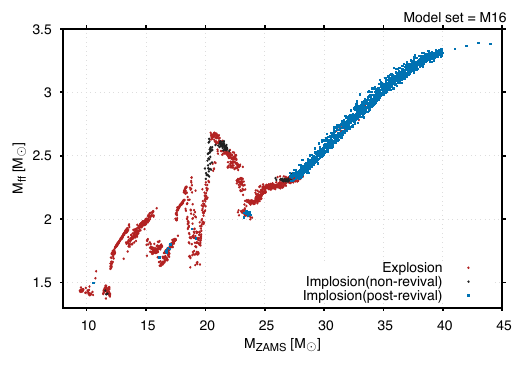}
    \includegraphics[width=0.45\linewidth]{ 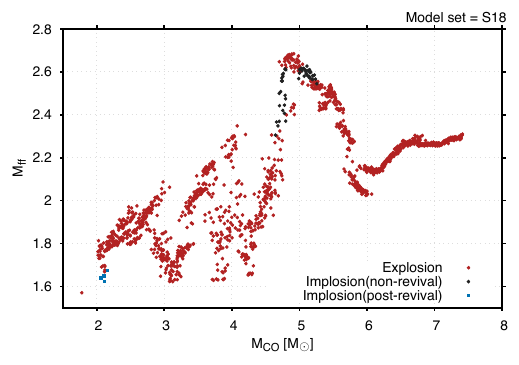}
    \includegraphics[width=0.45\linewidth]{ 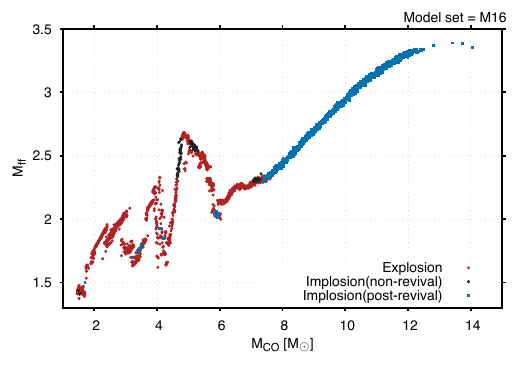}
    \includegraphics[width=0.45\linewidth]{ 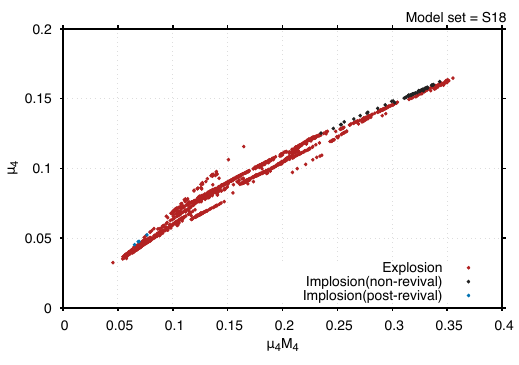}
    \includegraphics[width=0.45\linewidth]{ 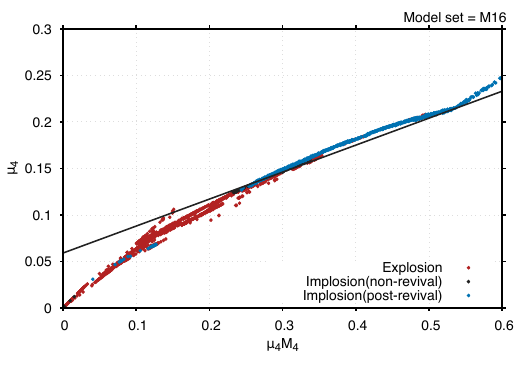}
    \end{center}
    \caption{Explodability map computed with the updated parameter set. Color coding is the same as in Figure~\ref{fig:structure-compactness}. \\
    Alt text: Six scatter plots showing explodability maps computed with the updated parameter set. The top row plots free-fall mass coordinate against ZAMS mass for S18 and M16, the middle row plots free-fall mass coordinate against CO core mass, and the bottom row plots $\mu_4$ against $\mu_4M_4$; red, black, and blue points mark explosion, non-revival implosion, and post-revival implosion.}
    \label{fig:results:structure variable}
\end{figure*}

Figure~\ref{fig:results:structure variable} shows the explodability distribution for the S18 and M16 progenitor sets as functions of $M_{\rm ff}$ and $\mu_4$--$M_4$. For S18, most failed explosions cluster near $\sim 12\,M_\odot$ and in the $\sim 19$--$22\,M_\odot$ range. For M16, failed explosions are more widely distributed, including $\sim 11$--$12\,M_\odot$, $\sim 16$--$17\,M_\odot$, around $\sim 19\,M_\odot$, $\sim 20$--$22\,M_\odot$, around $\sim 23\,M_\odot$, and above $\sim 26\,M_\odot$. The $\mu_4$--$M_4$ boundary for M16 is $\mu_4 = 0.29\,(\mu_4 M_4)+0.059$, with 183 / 2854 misclassified. The low-mass region cannot yield a boundary, and S18 similarly cannot derive a boundary due to the lack of a large-mass sample. These patterns differ from the distribution obtained with the original parameter set. The updated results therefore motivate a re-evaluation of how well structure-based variables predict explodability. We address this in Section~4.3. 

In the middle row of Figure~\ref{fig:results:structure variable}, we show explodability in the $M_{\rm ff}$--$M_{\rm CO}$ plane. Using $M_{\rm CO}$ reduces ambiguity when comparing progenitor sets with different microphysics. \citep[e.g., ][]{Patton2020, laplace2025} The distribution remains broad: even at similar $M_{\rm CO}$, models cover a range of $M_{\rm ff}$ and include both explosions and implosions, indicating that the separation is not captured by a single structural coordinate.

\section{Discussion}

\subsection{Implosion Samples}

We calibrate the model by minimizing a $\chi^2$-like mismatch metric and selecting the parameter set that yields the smallest mismatch to the 3D explosion outcomes. We then inspect the resulting solutions and choose a parameter set that provides a reasonable overall behavior among the tested candidates. The 3D dataset contains two direct-implosion cases at $12.25\,M_\odot$ and $14\,M_\odot$. Our best-fit metric does not include these implosion samples explicitly, so the optimization is not designed to reproduce them. Instead, after the fit we compare the predicted implosion regions with the 3D implosion samples and assess whether the model reproduces them qualitatively.

We also attempted to incorporate the explosion--implosion threshold implied by these two implosion samples when fitting $\alpha_{\rm turb}(M_{\rm ff})$. We found that reproducing both implosions requires extremely small values of $\alpha_{\rm turb}$. However, when we examined progenitor-structure variables, we did not identify any variable that could naturally explain such a low $\alpha_{\rm turb}$ in these cases. For this reason, the same progenitor that implodes in 3D does not necessarily implode in our optimized model. Nevertheless, several non-exploding models appear in the surrounding mass range, as shown in Figure~\ref{fig:results:structure variable}. This still provides a useful guide for locating mass ranges where explosions tend to fail and for exploring the physical reasons behind those failures.

In this low-mass regime, a similar tension also appears when applying the pre-SN criteria of \citet{maltsev2025}. The $12.25\,M_\odot$ and $14\,M_\odot$ cases fall below their lower compactness threshold and would therefore be assigned to the exploding class in that scheme. We summarize this point and the related model ranges in Appendix~\ref{app:maltsev}. This indicates that low-mass direct-implosion outcomes remain challenging for fast pre-SN classification schemes and may be sensitive to the adopted 3D dataset and modelling assumptions.

\subsection{Parameters}
The original parameter set was mainly calibrated to 2D simulations. Recent 3D studies suggest that multi-dimensional effects can make explosions easier than in 2D. Our updated parameters therefore shift the semi-analytic model toward more successful explosions overall.

The fitted relation $\alpha_{\rm turb}(M_{\rm ff})$ is a major source of uncertainty. The current 3D calibration sample is sparse in $M_{\rm ff}$, with large gaps, so the fitted curve is only loosely constrained (Figure~\ref{fig:optimization-fitaturb}). We tested several alternative fitting forms. Small changes in $\alpha_{\rm turb}$ can change the explodability classification, which highlights the model sensitivity to this parameter. We further assess the robustness of the inferred $\alpha_{\rm turb}(M_{\rm ff})$ trend by comparing with additional 3D outcomes (including \citealt{mueller2019}); the detailed discussion is provided in Appendix~\ref{app:3d-uncertainty}.

Figure~\ref{fig:discussion:const-aturb} illustrates the explodability distribution obtained when $\alpha_{\rm turb}$ is fixed to a single constant value for all progenitors. The three panels demonstrate that a constant-$\alpha_{\rm turb}$ assumption introduces strong trade-offs across the mass range.
In the first panel, we adopt the parameter set that minimizes $\bar{\chi}_{\rm tot}^2$ in our stepwise best-fit procedure (we first scan $\alpha_{\rm turb}$, $\zeta$, and $\tau_{1.5}$, and then scan $\alpha_{\rm out}$ and $\beta_{\rm expl}$). In this case, implosions concentrate mainly at low ZAMS masses, while most high-mass progenitors, especially those above $30\,M_\odot$, are classified as successful explosions. We consider this distribution unrealistic.
In the second panel, we fix all parameters except $\alpha_{\rm turb}$ to the optimized values listed in Table~\ref{tab:opt-values} and perform a one-dimensional best-fit scan over $\alpha_{\rm turb}$ only. Compared with our suggested (fiducial) results in Section~3, the main differences appear in the $20$--$22,M_\odot$ range and above $\sim 27,M_\odot$. In our fiducial calibration, many progenitors in these mass ranges are classified as failed explosions, whereas the constant-$\alpha_{\rm turb}$ assumption yields successful explosions for most of them. Because the current 3D calibration sample is small, we cannot determine which distribution is more accurate.
In addition, the constant-$\alpha_{\rm turb}$ case cannot reproduce the low-mass direct-implosion trend around $\sim 12\,M_\odot$ seen in 3D. If we reduce $\alpha_{\rm turb}$ to force implosions in this low-mass range, the outcome becomes the third panel in Figure~\ref{fig:discussion:const-aturb}: an excessively small $\alpha_{\rm turb}$ drives most progenitors to implode, which is also not desirable.

Figure~\ref{fig:discussion:fit-type2} shows one alternative fit. In that test, we keep the other four parameters fixed to the values in Table~\ref{tab:opt-values}. The alternative fit is written as 
\begin{equation}
    \alpha_{\rm turb}=
\begin{cases}
-1.26\,M_{\rm ff}^2 + 4.65\,M_{\rm ff} - 3.02, & M_{\rm ff}\le 2.105,\\
\ \ 0.05\,M_{\rm ff} + 1.08, & M_{\rm ff} > 2.105.
\end{cases}
\end{equation}
Compared with our fiducial results, the alternative fit produces fewer implosions above $20\,M_\odot$ and more implosions below $20\,M_\odot$.

We adopt the parameter set presented in Results as our fiducial choice. It provides the most consistent match to the available 3D outcome and reproduces the qualitative trends reported by \citet{burrows2024}. Nevertheless, the treatment of black-hole-forming cases remains a limitation. In the 3D sample, progenitors of $19.56\,M_\odot$ and $40\,M_\odot$ undergo shock revival and later collapse to black holes, while $12.25\,M_\odot$ and $14\,M_\odot$ collapse without shock revival. The M\"uller model can only classify the final remnant by checking whether the PNS mass exceeds the adopted maximum mass. It does not model the late-time accretion and fallback physics in a self-consistent way. This makes the post-revival black-hole channel difficult to constrain with the current framework. Improving this aspect likely requires revisiting the physical treatment in phase~II, where the model currently assumes negligible accretion.

The calibration would also benefit from additional 3D samples. More models, especially near the transition regions, would better constrain $\alpha_{\rm turb}(M_{\rm ff})$ and reduce the degeneracy among parameters. With a larger sample, the fitting could be made more robust and the resulting explodability maps would be less sensitive to the chosen functional form.

\subsection{Structure-based variables}

Our updated model predicts that explosions can occur for relatively compact progenitors in the $20$--$22\,M_\odot$ range. This behavior differs from many traditional 1D results, but it is closer to recent 3D outcome. As a result, single-parameter indicators such as the compactness $\xi_{2.5}$ and the free-fall mass coordinate $M_{\rm ff}$ have clear limitations as universal explodability criteria. In particular, they do not robustly separate exploding and non-exploding models below $\sim 30\,M_\odot$ in our sample.

At the same time, $\xi_{2.5}$ and $M_{\rm ff}$ remain useful for describing explosion outcome. As shown in Figure~\ref{fig:results:explosion characteristics}, explosion energy, ejected $^{56}$Ni mass, and PNS mass show approximately monotonic trends with $M_{\rm ff}$, much better than $M_{\rm ZAMS}$. These parameters are therefore better suited for predicting broad trends in explosion outcome than for providing a sharp success/failure boundary.

The two-parameter $\mu_4$--$M_4$ criterion retains more predictive power because it encodes the density gradient near the $s=4$ interface. However, its separating line depends on the available sample and on details of the progenitor set. With limited samples, the fitted boundary can shift noticeably. The criterion may therefore require re-fitting when applied to different progenitor sets or when new calibration data become available.

\begin{figure}
    \begin{center}
    \includegraphics[width=1\linewidth]{ 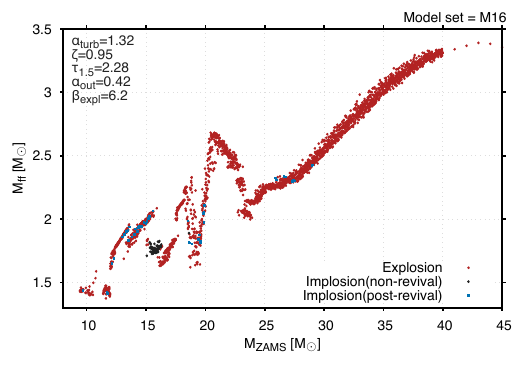}
    \includegraphics[width=1\linewidth]{ 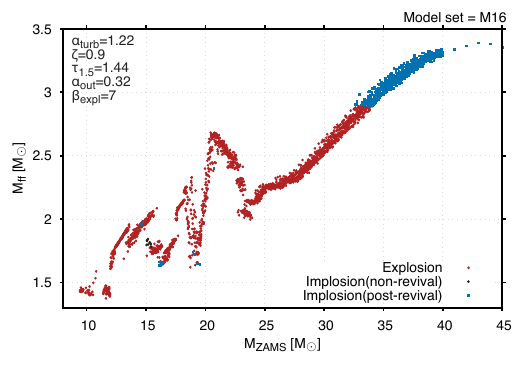}
    \includegraphics[width=1\linewidth]{ 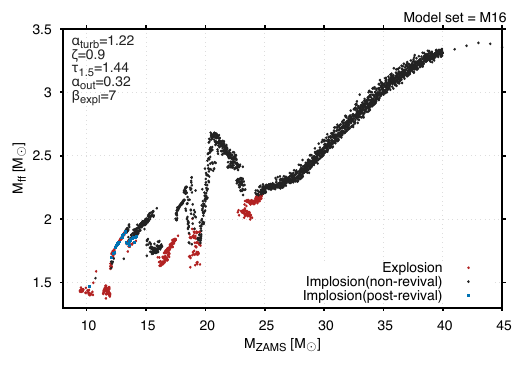}
    \end{center}
    \caption{Explodability maps obtained under the assumption of a constant $\alpha_{\rm turb}$. Color coding is the same as in Figure~\ref{fig:structure-compactness}. \\
    Alt text: Three vertically stacked explodability maps for M16 under different constant $\alpha_{\rm turb}$ assumptions. Red, black, and blue points show that changing a single constant $\alpha_{\rm turb}$ shifts the distribution of successful explosions and implosions across the mass range.}
    \label{fig:discussion:const-aturb}
\end{figure}

\begin{figure}
    \begin{center}
    \includegraphics[width=1\linewidth]{ 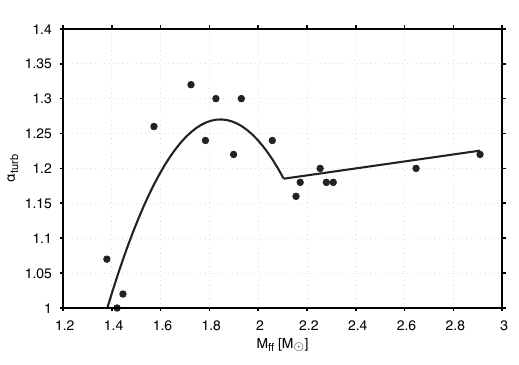}
    \includegraphics[width=1\linewidth]{ 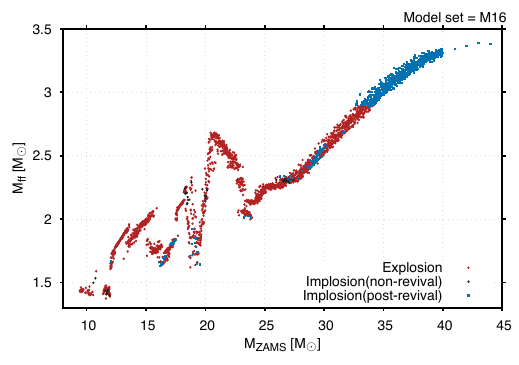}
    \end{center}
    \caption{Alternative piecewise fit for $\alpha_{\rm turb}(M_{\rm ff})$. And explodability map computed with the alternative updated parameter set. Color coding is the same as in Figure~2. \\
Alt text: A two-panel figure showing an alternative fit for $\alpha_{\rm turb}$ as a function of free-fall mass coordinate and the resulting M16 explodability map. The upper panel shows black points with a fitted curve, and the lower panels show that the alternative fit changes where implosions occur across the mass range.}
    \label{fig:discussion:fit-type2}
\end{figure}

\begin{figure*}
    \begin{center}
    \includegraphics[width=0.45\linewidth]{ 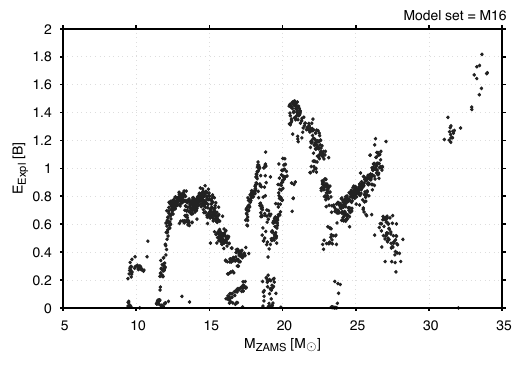}
    \includegraphics[width=0.45\linewidth]{ 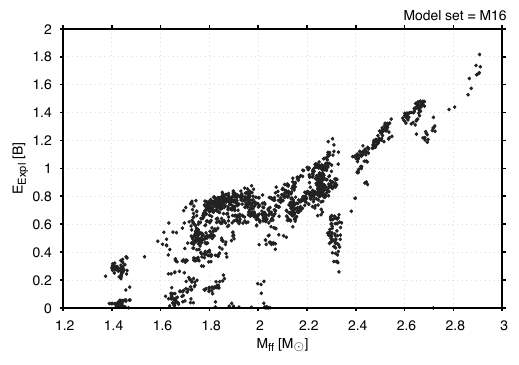}
    \includegraphics[width=0.45\linewidth]{ 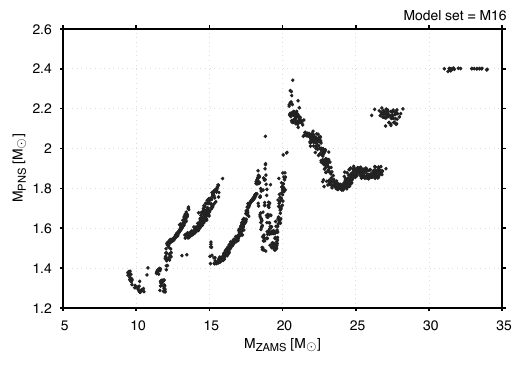}
    \includegraphics[width=0.45\linewidth]{ 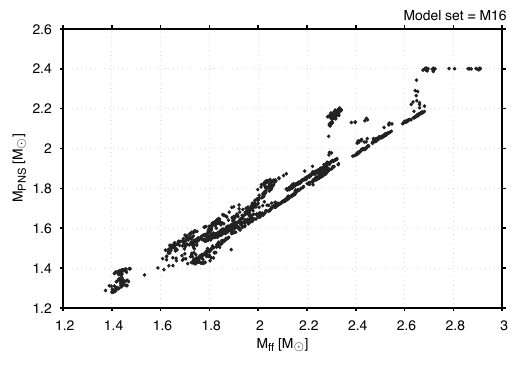}

    \includegraphics[width=0.45\linewidth]{ 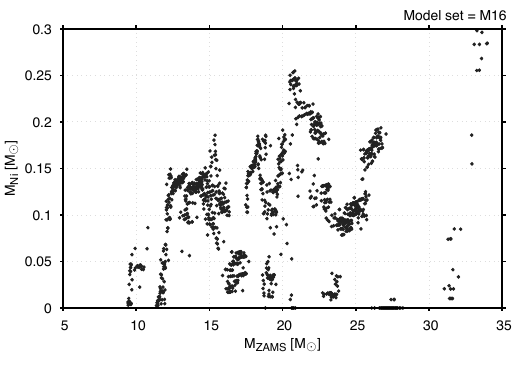}
    \includegraphics[width=0.45\linewidth]{ 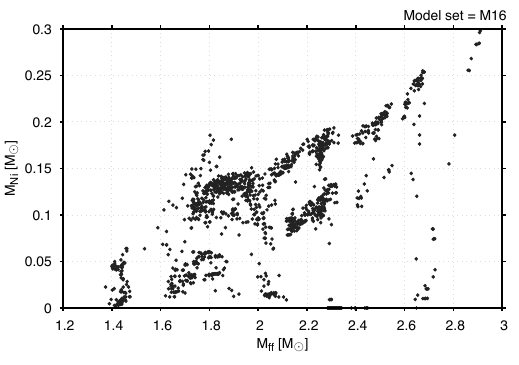}
    \end{center}
    \caption{Explosion characteristics as functions of $M_{\rm ZAMS}$ (left) and $M_{\rm ff}$ (right).\\
    Alt text: Six scatter plots arranged in three rows and two columns for M16. Explosion energy, proto-neutron-star mass, and nickel mass are plotted against ZAMS mass on the left and free-fall mass coordinate on the right; the relations are more nearly monotonic with free-fall mass coordinate.}
    \label{fig:results:explosion characteristics}
\end{figure*}

\section{Conclusion}
 
This paper aims to answer the question of how a fast one-dimensional (1D) framework can better track the trends seen in an adopted set of three-dimensional (3D) neutrino-driven explosions. 
Using a parameter-tunable semi-analytic 1D explosion model, we compare against one representative 3D calibration set and systematically evaluate the validity and limitations of commonly used structure-based criteria in this context. We stress, however, that the present study is calibrated mainly to this single 3D dataset, and that current 3D explosion outcomes themselves still carry non-negligible theoretical uncertainty.

In terms of motivation, 3D simulations can represent multi-dimensional effects such as neutrino heating, convection, and SASI more realistically, but their computational cost is high, which makes systematic scans over large progenitor samples difficult. To address this, we adopt and reproduce the full workflow of the semi-analytic model, and we use an implementation that incorporates corrections from subsequent work to ensure methodological reproducibility and consistent cross-comparisons. For progenitors, we mainly use the Sukhbold progenitor suites (S16 and S18) and the progenitor set used by M\"uller (M16) as our reference samples.

Methodologically, we separate explosion outcome into two types of information: (i) explodability (whether the model explodes), and (ii) explosion outcome and remnant properties (explosion energy, PNS mass, and $^{56}$Ni mass). To bring the 1D results closer to the 3D trends, we define a $\chi^2$-like mismatch metric based on these three outcome quantities and optimize the model parameters within their allowed ranges. In the optimization, we first focus on $\alpha_{\rm turb}$ and fit it as a function of $M_{\rm ff}$. We then scan $\zeta$, $\tau_{1.5}$, $\beta_{\rm expl}$, and $\alpha_{\rm out}$ to obtain an updated parameter set.

With the updated parameters, the overall agreement between the 1D model and the 3D results improves. The improvement is especially clear for low-mass progenitors ($M<10,M_\odot$), where the differences in $E_{\rm expl}$, $M_{\rm PNS}$, and $M_{\rm Ni}$ relative to the 3D values shrink substantially. The updated set produces an explodability distribution that is closer to the trend suggested by the 3D sample, and it changes the explosion classification at several key masses. At the same time, we show that explodability is sensitive to the fitted form of $\alpha_{\rm turb}(M_{\rm ff})$. With a sparse calibration sample, different fits for $\alpha_{\rm turb}$ can shift the failure distribution between high- and low-mass regimes, indicating that the calibration is not yet unique under the current constraints.

For structure-based criteria, we compute and compare $\xi_{2.5}$, $M_{\rm ff}$, and $\mu_4$--$M_4$. We find that single-parameter indicators have limited ability to provide a universal success/failure boundary, especially in the intermediate-mass regime where exploding and non-exploding models overlap. However, they describe trends in explosion properties (energy, PNS mass, and Ni yield) more stably and show approximately monotonic relations. They are therefore better suited for characterizing statistical trends in explosion outcome than for serving as universal explodability criteria. In contrast, the two-parameter $\mu_4$--$M_4$ criterion can, to some extent, separate a failing branch that deviates from the main trend, but its boundary depends more strongly on sample size and on the chosen progenitor set. It often needs to be re-calibrated when applied across datasets.

Finally, we highlight the main limitations of the current framework and directions for improvement. The semi-analytic model still cannot reliably constrain the channel in which shock revival is followed by late-time collapse to a black hole. This is mainly because late-time accretion and the associated physics are strongly simplified in the model. A richer set of 3D samples-especially near transition masses and along black-hole formation channels-would better constrain $\alpha_{\rm turb}(M_{\rm ff})$, reduce parameter degeneracies, and support a more robust parameter-update strategy.

Overall, this work shows that a parameter-tunable semi-analytic 1D model can absorb trends revealed by 3D simulations while retaining high computational efficiency. Beyond enabling large-sample explodability surveys and re-evaluations of structure-based criteria, the calibrated model also has practical value for yield-based applications. A substantial body of 1D nucleosynthesis calculations still relies on artificial explosion prescriptions to produce yield tables over wide progenitor grids. These yields are widely used in galactic chemical evolution and in comparisons with abundance observations. By providing fast, physically motivated estimates of explodability, $E_{\rm expl}$, and $^{56}$Ni production, our approach can offer calibration targets to tune such artificial-explosion prescriptions. This can improve the consistency of yield tables and strengthen related studies, including progenitor inference from supernova light curves that still largely depend on 1D explosion models.

\begin{ack}

This work was partly supported by JSPS KAKENHI Grant Numbers JP21H01123.

We thank Andr\'es Gonz\'alez, Bernhard M\"uller, Jiabao Liu, Kehan Zou and Koh Nakamura for helpful discussions.

\end{ack}

{
\appendix

\section{Uncertainties in the 3D calibration and the broader landscape of 3D efforts}
\label{app:3d-uncertainty}

Our baseline calibration uses the 3D neutrino-driven explosion outcomes reported by \citet{burrows2024}. As stated in the main text, we treat this dataset as a practically useful calibration set rather than as a definitive ground truth. Independent state-of-the-art 3D simulations can yield different outcomes even for similar progenitors. Such tensions likely reflect differences in neutrino-transport treatments, gravity approximations, microphysics, numerical resolution, and energy conservation, among other implementation choices. For this reason, we interpret our optimized parameters as a calibration to a representative set of 3D outcomes, and we do not treat any single 3D dataset as absolutely correct. We also caution that some individual outcomes in current 3D simulations (e.g., black-hole formation for progenitors near $\sim 12\,M_\odot$) may be in tension with observational inferences for exploding stars in a similar mass range, and should therefore not be over-interpreted as definitive.

To place our calibration in a broader context, we also consider a small number of independent 3D neutrino-driven simulations from other efforts. Besides the FORNAX calculations discussed in this work, widely used self-consistent 3D approaches in the literature include CoCoNuT-FMT and PROMETHEUS-VERTEX. Because published 3D studies sometimes employ parameterized extensions to follow the evolution to later times, we restrict our supplementary comparison, as far as can be judged from the published descriptions, to models that are (i) solar metallicity, (ii) non-rotating, and (iii) evolved as self-consistently as possible during the explosion phase. These restrictions reduce ambiguity, but they also significantly reduce the number of usable models. We therefore do not use these additional cases for the primary parameter optimization. Instead, we use them as a consistency check on the optimized model.

The fitted dependence $\alpha_{\rm turb}(M_{\rm ff})$ is a key ingredient in our calibration, and it is affected by the sparse sampling of the available 3D dataset. As an additional check, we include best-fit $\alpha_{\rm turb}$ values inferred from the four 3D cases of \citet{mueller2019} (orange points in Figure~\ref{fig:app_aturb}). These points are obtained by fixing the remaining parameters to the optimized set and scanning $\alpha_{\rm turb}$ to minimize the same error metric used in the primary calibration, using the reported 3D explosion properties for each case. The revised fit is slightly shifted around $M_{\rm ff} \sim 2.0$--$2.5\,M_\odot$, which in turn modifies the explodability distribution in the lower panel of Figure~\ref{fig:app_aturb}. One clear outlier in the top panel of Figure~13 appears near $M_{\rm ff}=1.8\,M_\odot$ and corresponds to the $12.5\,M_\odot$ model from \citet{mueller2019}. In our semi-analytic comparison, this model cannot be reproduced very well by a simple adjustment of the Müller-model parameters. We therefore do not regard the corresponding best-fit $\alpha_{\rm turb}$ value as strong evidence for a real sharp feature at this $M_{\rm ff}$. Instead, it suggests that some progenitors in this mass range are difficult to describe with a simple smooth $\alpha_{\rm turb}(M_{\rm ff})$ relation, especially because the current 3D sample is still limited. We discuss this point further in Section~\ref{app:maltsev}. Overall, this provides an independent consistency check, although the sample is still too small to uniquely determine the functional form.

For reference, the revised $\alpha_{\rm turb}(M_{\rm ff})$ relation used in Figure~\ref{fig:app_aturb} is the following piecewise function:
\begin{equation}
\alpha_{\rm turb}=
\begin{cases}
-1.795\,M_{\rm ff}^2 + 6.403\,M_{\rm ff} - 4.419, & M_{\rm ff}\le 1.93,\\
\ \ 0.871\,M_{\rm ff}^2 - 3.893\,M_{\rm ff} + 5.521, & 1.93 < M_{\rm ff}\le 2.28,\\
\ \ 0.076\,M_{\rm ff} + 0.998, & M_{\rm ff} > 2.28.
\end{cases}
\end{equation}

\begin{figure}
    \begin{center}
    \includegraphics[width=1\linewidth]{ 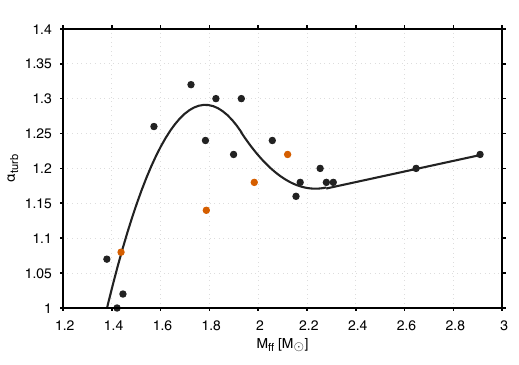}
    \includegraphics[width=1\linewidth]{ 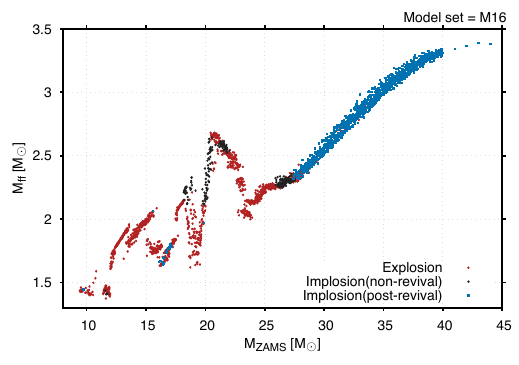}
    \end{center}
    \caption{Same as Figure~7, but with additional best-fit $\alpha_{\rm turb}$ values inferred from the four 3D cases of \citet{mueller2019} included in the fit. Top: best-fit $\alpha_{\rm turb}$ values (black points) and the updated fitted relation $\alpha_{\rm turb}(M_{\rm ff})$ (black curve); orange points highlight the additional best-fit values from \citet{mueller2019}. Bottom: explodability map for the M16 progenitor set computed with the updated parameter set based on this revised fit. Color coding is the same as in Figure~2.}
    \label{fig:app_aturb}
    \par\smallskip
    \noindent\textbf{Alt text:} A two-panel figure extending the $\alpha_{\rm turb}$--$M_{\rm ff}$ fit by adding best-fit points from \citet{mueller2019} and showing the impact on explodability for the M16 progenitors. The top panel shows black points and a smooth fitted curve, with additional orange points. The bottom panel shows the $M_{\rm ff}$--$M_{\rm ZAMS}$ explodability map; the revised fit shifts where non-exploding regions appear.
\end{figure}

To illustrate the current spread among published 3D outcomes, we note a few examples at broadly similar ZAMS masses. \citet{burrows2024} report $E_{\rm expl}\approx 0.325\,{\rm B}$ for an $\sim 11\,M_\odot$ progenitor, while \citet{mueller2019} obtain $E_{\rm expl}=0.199\,{\rm B}$ for $11.8\,M_\odot$. For $\sim 15\,M_\odot$, \citet{burrows2024} find $E_{\rm expl}\sim 0.3\,{\rm B}$, whereas \citet{mueller2020} report $E_{\rm expl}=0.023\,{\rm B}$ for their $15\,M_\odot$ model. For $\sim 18\,M_\odot$, \citet{burrows2024} report $E_{\rm expl}\sim 0.55\,{\rm B}$ and \citet{mueller2017} obtain $E_{\rm expl}\sim 0.77\,{\rm B}$. Given differences in progenitor structures, diagnostics, and numerical setups, and given the limited number of directly comparable models, we do not view these examples as establishing a clear systematic contradiction. Instead, they highlight the current theoretical uncertainty and motivate caution when calibrating against any single 3D dataset.

We report the corresponding semi-analytic predictions using two parameter choices. The results computed with the fiducial parameter set adopted in the main text are listed in Table~\ref{tab:muller3D}. The results computed with the revised $\alpha_{\rm turb}(M_{\rm ff})$ fit shown in Figure~\ref{fig:app_aturb}, while keeping the remaining parameters fixed to the fiducial set, are listed in Table~\ref{tab:muller3D_opti}.

\begin{table*}
  \tbl{Comparison with other 3D results (fiducial parameter set).\footnotemark[*]}{%
  \begin{tabular}{cccccccc}
      \hline
      $M_{\rm ZAMS}$[$M_\odot$] & Reference & $E_{\rm Expl,3D}$[B] & $E_{\rm Expl,1D}$[B] & $M_{\rm PNS,3D}$[$M_\odot$] & $M_{\rm PNS,1D}$[$M_\odot$] & $M_{\rm Ni,3D}$[$M_\odot$] & $M_{\rm Ni,1D}$[$M_\odot$]  \\ 
      \hline
      11.8 & \cite{mueller2019} & 0.199 & 0.234 & 1.35 & 1.34 & 0.024 & 0.018 \\
      12.5 & \cite{mueller2019} & 0.156 & 0.761 & 1.61 & 1.54 & 0.013 & 0.130 \\
      15   & \cite{mueller2020} & 0.023 & 0.687 & -    & 1.73 & -     & 0.106 \\
      18   & \cite{mueller2017} & 0.770 & 0.797 & 1.87 & 1.75 & -     & 0.151 \\
      \hline
    \end{tabular}}\label{tab:muller3D}
\begin{tabnote}
\footnotemark[*] For a uniform comparison, the 1D values are computed with our implementation using the M16 progenitor profiles wherever possible (see text).
\end{tabnote}
\end{table*}

\begin{table}
  \tbl{Comparison with other 3D results (revised $\alpha_{\rm turb}(M_{\rm ff})$). \footnotemark[*]}{%
  \begin{tabular}{cccc}
      \hline
      $M_{\rm ZAMS}$[$M_\odot$] & $E_{\rm Expl,1D}$[B] & $M_{\rm PNS,1D}$[$M_\odot$] & $M_{\rm Ni,1D}$[$M_\odot$]  \\ 
      \hline
      11.8 & 0.015 & 1.34 & 0.009 \\
      12.5 & 0.749 & 1.54 & 0.130 \\
      15   & 0.571 & 1.76 & 0.056 \\
      18   & 0.800 & 1.76 & 0.148 \\
      \hline
    \end{tabular}}\label{tab:muller3D_opti}
\begin{tabnote}
\footnotemark[*] Same as Table~\ref{tab:muller3D}, but the 1D values are computed using the revised $\alpha_{\rm turb}(M_{\rm ff})$ fit shown in Figure~\ref{fig:app_aturb}, while keeping the remaining parameters fixed to the fiducial set.
\end{tabnote}
\end{table}

\section{An explicit comparison to Maltsev et al.\ (2025)}
\label{app:maltsev}

Several studies closely related to \citet{muller2016} have explored how the semi-analytic framework can be tuned to match specific observational or theoretical constraints. For example, \citet{Schneider2021} adjusted the semi-analytic parameters to $\beta_{\rm expl}=3.3$ and $\alpha_{\rm out}=1.22$ and adopted a maximum gravitational neutron-star mass of $M_{\rm NS,grav}^{\max}=2.0\,M_\odot$. Their goal was to place the mean explosion energy of Type~IIP supernovae in the observed range of $\sim 0.5$--$1.0\,{\rm B}$, while retaining a high-energy tail that allows single-star progenitors to reach explosion energies up to $\sim 3\,{\rm B}$.

\citet{maltsev2025} adopted a complementary perspective. They started from explodability predictions associated with \cite{muller2016} and compressed these predictions into a multi-parameter pre-SN criterion based on a small number of scalar structural quantities. Their prescription first checks whether $\xi_{2.5}$, $s_{\rm c}$, and $M_{\rm CO}$ fall below lower thresholds (classified as successful) or above upper thresholds (classified as failed). The quoted lower/upper thresholds are: 
\begin{align}
\xi_{2.5} &= 0.314/0.544, \\
s_{\rm c}/(N_A k_B)&=0.988/1.169, \\
M_{\rm CO}/M_\odot &= 5.6/16.2, \\
\mu_4 M_4 &= 0.247/0.421. 
\end{align}
If any of $\xi_{2.5}$, $s_{\rm c}$, or $M_{\rm CO}$ lies below the corresponding lower threshold, the model is classified as a successful SN; if any lies above the corresponding upper threshold, it is classified as a failed SN. For models in the intermediate overlap region, the final fate is determined by a linear boundary in the $(\mu_4 M_4,\mu_4)$ plane,
\begin{align}
\mu_4 < 0.005 + 0.420\,(\mu_4 M_4),
\end{align}
where models satisfying this condition are classified as failed SNe and the remainder as successful SNe.

We view this approach as complementary to ours. Our objective is not only to classify explodability, but also to reproduce trends in explosion properties (e.g., $E_{\rm expl}$, $M_{\rm PNS}$, and $^{56}$Ni yields) within a fast neutrino-driven framework calibrated to 3D outcomes. The performance of any structure-based criterion is inevitably tied to the specific progenitor sets and 3D datasets used for calibration, and should therefore be interpreted with that dependence in mind.

In their study, the authors applied their pre-SN criteria to 29 3D outcomes from the Monash group (CoCoNuT-FMT) and the Garching group (PROMETHEUS-VERTEX). Their prescription correctly classifies 25 models but fails for four cases (m39, z85, z40, and s14). In particular, s14 is classified as a non-explosion in 3D, while their boundary predicts an explosion. Tensions of this type also appear in our comparison. The $12.25\,M_\odot$ and $14\,M_\odot$ progenitors in the S18 subset of \citet{burrows2024} are classified as non-explosions in 3D but as explosions in our optimized model, and we also do not obtain a satisfactory reproduction for the $12.5\,M_\odot$ progenitor from \citet{mueller2019} (Section~\ref{app:3d-uncertainty}). For these two low-mass S18 cases, the compactness values are $\xi_{2.5}=0.033$ and $0.120$, respectively. Applying the pre-SN threshold rules of \citet{maltsev2025} to these values also yields a classification of successful explosions, because both lie well below the lower compactness threshold. This suggests that low-mass direct-implosion outcomes can be difficult to capture with threshold-based pre-SN criteria and may be sensitive to the adopted 3D dataset and modelling assumptions. A more definitive assessment for this mass range will require a larger set of published 3D samples.
}

\bibliographystyle{aasjournal}
\bibliography{reference}{}

\end{document}